\documentclass[12pt]{book}

\usepackage[dvips]{graphicx,color}
\usepackage{makeidx,tsukuba}
\newcommand{\U}[1]{\ensuremath{\mathrm{\ #1}}}

\makeauthorindex
\makeindex

\begin{document}

\BookTitle{\itshape The 28th International Cosmic Ray Conference}
\CopyRight{\copyright 2003 by Universal Academy Press, Inc.}
\pagenumbering{arabic}

\chapter{
Whipple observations of 1ES1959+650: an Update.}

\author{%
%
%
J.~Holder$^{1,2}$, I.H.~Bond, P.J.~Boyle, S.M.~Bradbury, J.H.~Buckley,
D.~Carter-Lewis, O.~Celik, W.~Cui, M.~Daniel, M.~D'Vali,
I.de~la~Calle~Perez, C.~Duke, A.~Falcone, D.J.~Fegan, S.J.~Fegan,
J.P.~Finley, L.F.~Fortson, J.~Gaidos, S.~Gammell, K.~Gibbs,
G.H.~Gillanders, J.~Grube, J.~Hall, T.A.~Hall, D.~Hanna, A.M.~Hillas,
D.~Horan, A.~Jarvis, M.~Jordan, G.E.~Kenny, M.~Kertzman,
D.~Kieda, J.~Kildea, J.~Knapp, K.~Kosack, H.~Krawczynski, F.~Krennrich,
M.J.~Lang, S.~LeBohec, E.~Linton, J.~Lloyd-Evans, A.~Milovanovic,
P.~Moriarty, D.~Muller, T.~Nagai, S.~Nolan, R.A.~Ong, R.~Pallassini,
D.~Petry, B.~Power-Mooney, J.~Quinn, M.~Quinn, K.~Ragan, P.~Rebillot,
P.T.~Reynolds, H.J.~Rose, M.~Schroedter, G.~Sembroski, S.P.~Swordy,
A.~Syson, V.V.~Vassiliev, S.P.~Wakely, G.~Walker, T.C.~Weekes,
J.~Zweerink \\
{\it
(1) Department of Physics and Astronomy, University of Leeds, U.K.\\
(2) The VERITAS Collaboration--see S.P.Wakely's paper} ``The VERITAS
Prototype'' {\it from these proceedings for affiliations}
}

\section*{Abstract}
Strong flares of TeV gamma-ray emission up to a level of $\sim 5\U{Crab}$ were
detected by the Whipple 10~m atmospheric \v{C}erenkov telescope from the BL
Lacertae object 1ES1959+650 during May - July 2002. We report here the results
of follow up observations during 2002 - 2003.

\section{Introduction}
1ES1959+650 is a high frequency peak BL Lac object (HBL) at a redshift
$z=0.048$. It was first suggested as a candidate for TeV emission by Stecker,
de Jager and Salamon [10] and noted more recently by Costamante and Ghisellini
[1]. Initial weak detections by the Utah Seven Telescope Array [8] and HEGRA
instruments [5] made this object a prime candidate for ground-based
\v{C}erenkov telescopes during 2002 and observations during May - July were
rewarded with the detection of a period of strong TeV flare activity, reaching
a flux level of $\sim 5\U{Crab}$ [3,4]. This detection also triggered
contemporaneous target of opportunity measurements at $\sim10\U{keV}$ by the
pointed instruments on board RXTE [6] as well as observations at other
wavelengths [9]. Observations with the Whipple 10~m telescope are halted
during the summer months due to adverse weather conditions. In this paper we
summarise the Whipple observations of 1ES1959+650 including previously
unreported observations since September 2002.

\section{Instrument Status}
The configuration of the Whipple 10~m gamma-ray telescope is described in
detail in [2]. Briefly, the telescope consists of a 10~m reflector and a 490
pixel photomultiplier tube (PMT) camera. For the analysis presented here only
the high resolution ($0.12^{\circ}$ spacing) central 379 PMT pixels have been
used. The larger, outer 111 pixels were removed from the camera in January
2003. Using standard \textit{Supercuts} criteria to select gamma-ray events
and reject the background of hadronic cosmic rays on the basis of the shape of
the image in the camera, allows us to detect the Crab Nebula with a statistical
significance of $6\U{\sigma}$ in $1\U{hour}$.

\section{Observations and Analysis}
\begin{figure}[t]
  \begin{center}
    \includegraphics[height=12.0pc]{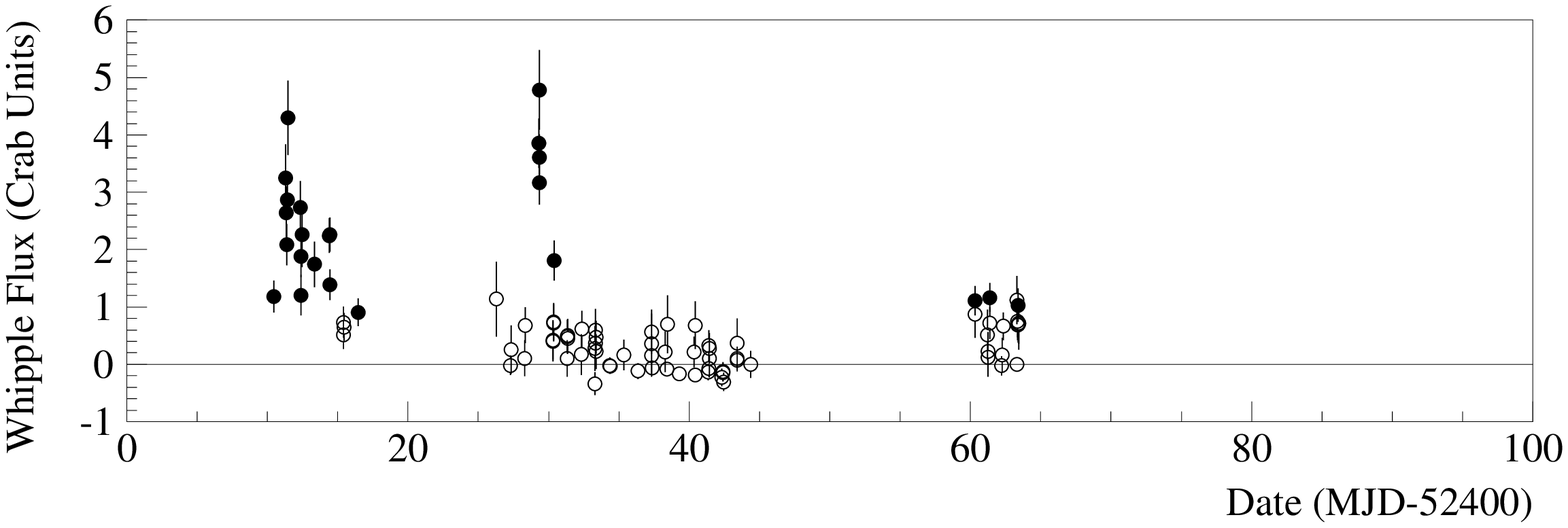}
    \includegraphics[height=12.0pc]{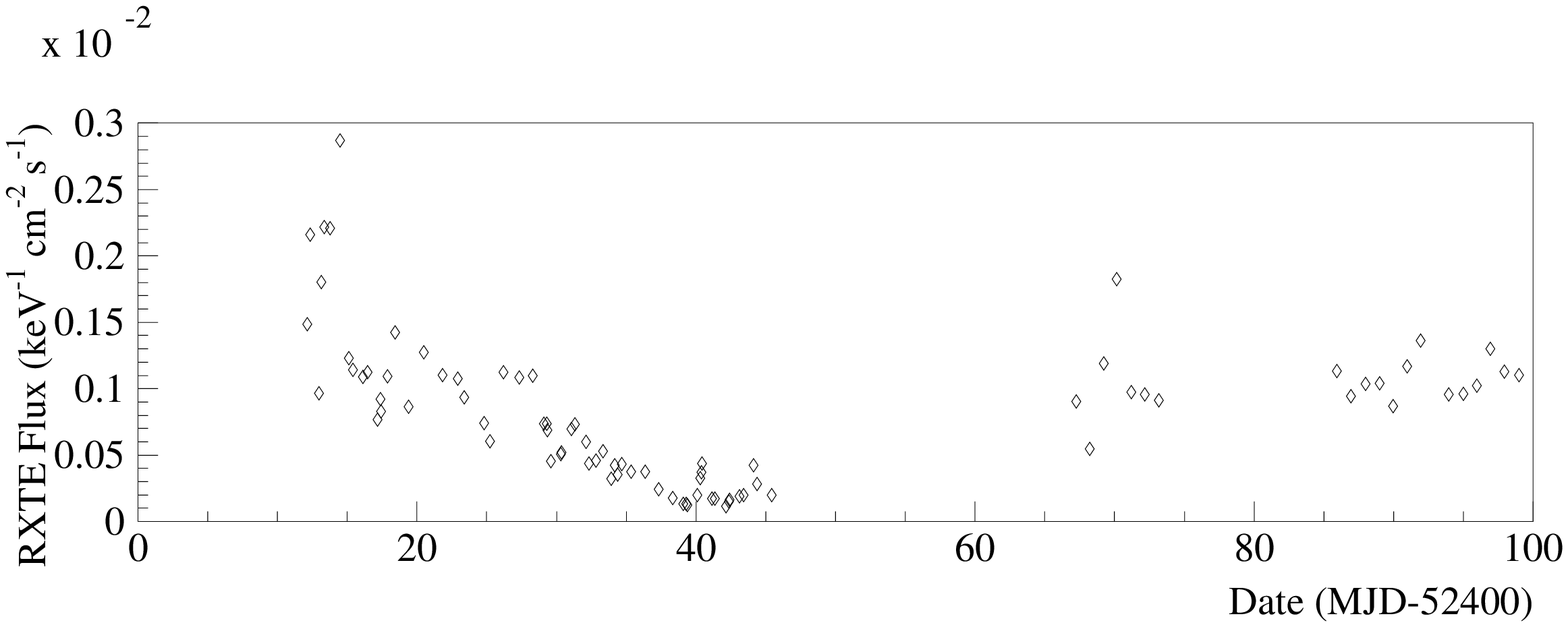} 
  \end{center}
  \vspace{-1.5pc}
  \caption{ The Whipple (top) and RXTE (bottom) light curves
  for 1ES1959+650 in May-July 2002. The filled Whipple points correspond to
  $>3\U{\sigma}$ detections. The RXTE data are from [6].}
\end{figure}

\begin{figure}[t]
  \begin{center}
    \includegraphics[height=13.5pc]{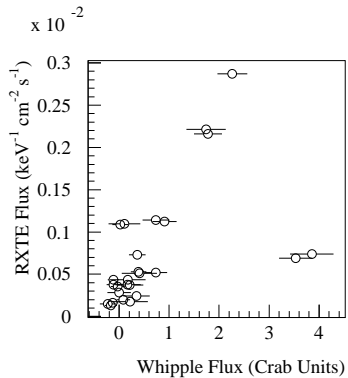} 
  \end{center}
  \vspace{-1.5pc}
  \caption{ The Whipple and RXTE fluxes. Only observations which overlap in
  time have been used.}
\end{figure}

Figure 1 shows the detailed light curves for the May - July 2002 observations
for both Whipple and RXTE data. The Whipple fluxes have been corrected for
source elevation and atmospheric changes using the method of LeBohec and
Holder [7]. A flare is visible at both wavelengths around MJD 52410 - 15,
however on MJD 52429 (June 4th) there was a clear example of a TeV
gamma-ray flare with no X-ray counterpart [6]. This flare corresponds to the
most rapid flux change in gamma-rays, with a doubling timescale of
$7\U{hours}$. This is further illustrated in Figure 2, where the X-ray flux is
plotted against the gamma-ray flux for all observations where there was
overlapping coverage. The ``orphan'' gamma-ray flare is in the bottom right
of the plot.

Figure 3 shows the daily average light curve for all Whipple observations of
1ES1959+650, including $11.2\U{hours}$ of observations from September
2002 to May 2003. The source has been relatively quiescent since September
2002, with only one day's observations (MJD 52584) revealing a significant
signal at $>3\U{\sigma}$.

\begin{figure}[t]
  \begin{center}
    \includegraphics[height=12.0pc]{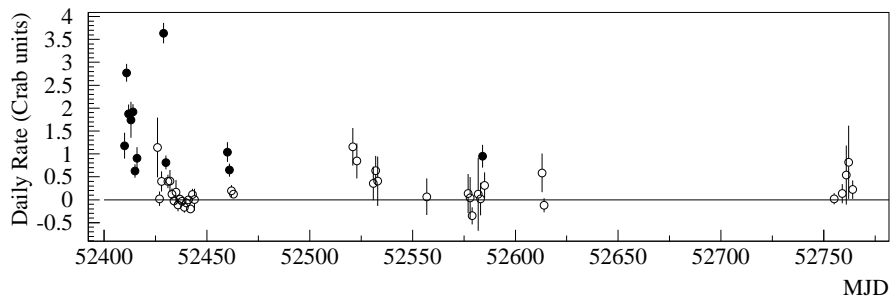} 
  \end{center}
  \vspace{-1.5pc}
  \caption{ The daily averaged gamma-ray flux for all Whipple
  observations. Filled points correspond to $>3\U{\sigma}$ detections. }
\end{figure}

\section{Discussion}
Figure 4 shows the $2-10\U{keV}$ flux from 1ES1959+650 as measured by the All
Sky Monitor (ASM) on board RXTE, averaged over bins of two weeks. The most
active (and variable) period occured in 2002 when the bright gamma-ray
flares were observed. Observations with the Whipple telescope are ongoing and
full results for 2003 will be reported at the conference. The intriguing
detection of a bright gamma-ray flare without an X-ray counterpart is
difficult to explain under the currently favoured one-zone synchrotron
self-Compton models for high energy emission from BL Lacs. Further
observations and detailed spectral analysis may help to clarify the
picture.

\begin{figure}[t]
  \begin{center}
    \includegraphics[height=12.0pc]{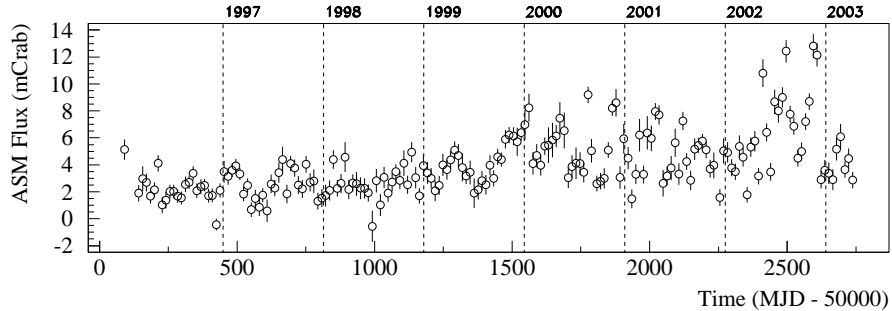} 
  \end{center}
  \vspace{-1.5pc}
  \caption{ The $2-10\U{keV}$ flux from 1ES1959+650 as measured by
  ASM in two week bins (quick-look results provided by the ASM/RXTE team).}
\end{figure}

\section{Acknowledgements}
We acknowledge the technical assistance of E. Roache and J. Melnick. This
research is supported by grants from the U.S. Department of Energy, by
Enterprise Ireland and by PPARC in the U.K.
\section{References}
\re
1.\ Costamante L.\& Ghisellini G.\ 2002, A\&A 384, 56
\re
2.\ Finley J.P.\ et al.\ 2001, in Proc. 27th ICRC,\ (Hamburg)
\re
3.\ Holder J.\ et al.\ 2003, ApJ 583, L9
\re
4.\ Horns D.\ et al.\ 2002, in High Energy Blazar Astronomy (Turku, Finland)
\re
5.\ Konopelko A.\ et al.\ 2002, April APS/HEAD Meeting,(Albuquerque)
\re
6.\ Krawczynski H.\ et al.\ in preparation.
\re
7.\ LeBohec S. \& Holder J.\ 2003, Astropart. Phys. 19, 221
\re
8.\ Nishiyama T.\ et al.\ 1999, in Proc. 26th ICRC\ (Salt Lake City)
\re
9.\ Schroedter M.\ et al.\ 2002, in The Universe Viewed in Gamma Rays
(Kashiwa, Japan)
\re
10.\ Stecker F.W., de Jager O.C. \& Salamon M.H.\ 1996, ApJ 473, 75

\endofpaper
\end{document}